\documentclass[preprint, 11pt, authoryear]{elsarticle}

\usepackage[T1]{fontenc}
\usepackage{microtype}
\usepackage{geometry}
\usepackage{booktabs}
\usepackage{makecell}
\usepackage{caption}
\usepackage[hyphens]{url}
\usepackage[hidelinks]{hyperref}
\hypersetup{breaklinks=true}
\journal{International Journal of Automotive Engineering (IJAE)}

\begin{document}

\begin{frontmatter}

% PAPER TITLE
\title{ndsintxn: An R Package for Extracting Information from Naturalistic Driving Data to Support Driver Behavior Analyses at Intersections}

% AUTHOR NAMES
\author[inst1]{Ashirwad Barnwal\corref{cor1}}
\ead{ashirwad@iastate.edu}
\cortext[cor1]{Corresponding author}
\author[inst2]{Jennifer Merickel}
\author[inst2]{Matthew Rizzo}
\author[inst3]{Luis Riera-Garcia}
\author[inst3]{Soumik Sarkar}
\author[inst1]{Anuj Sharma}

% AFFILIATION
\affiliation[inst1]{
    organization = {Department of Civil, Construction and Environmental Engineering, Iowa State University},
    city = {Ames},
    postcode = {50011}, 
    state = {Iowa},
    country = {United States}
}
\affiliation[inst2]{
    organization = {Department of Neurological Sciences, University of Nebraska Medical Center},
    city = {Omaha},
    postcode = {68198}, 
    state = {Nebraska},
    country = {United States}
}
\affiliation[inst3]{
    organization = {Department of Mechanical Engineering, Iowa State University},
    city = {Ames},
    postcode = {50010}, 
    state = {Iowa},
    country = {United States}
}

% ABSTRACT
\begin{abstract}
The use of naturalistic driving studies (NDSs) for driver behavior research has skyrocketed over the past two decades. Intersections are a key target for traffic safety, with up to 25-percent of fatalities and 50-percent injuries from traffic crashes in the United States occurring at intersections. NDSs are increasingly being used to assess driver behavior at intersections and devise strategies to improve intersection safety. A common challenge in NDS intersection research is the need for to combine spatial locations of driver-visited intersections with concurrent video clips of driver trajectories at intersections to extract analysis variables. The intersection identification and driver trajectory video clip extraction process are generally complex and repetitive. We developed a novel R package called ndsintxn to streamline this process and automate best practices to minimize computational time, cost, and manual labor. This paper provides details on the methods and illustrative examples used in the ndsintxn R package.
\end{abstract}

% KEYWORDS

\end{frontmatter}

% INTRODUCTION
\section{Introduction}
In today's data-driven world, naturalistic driving studies (NDSs) have become the gold standard for studying driver behavior under real-world traffic scenarios \citep{singh_analyzing_2021}. The NDS research method was first pioneered by the Virginia Tech Transportation Institute (VTTI) in the early 2000s as a way to overcome the limitations of retrospective and prospective studies (e.g., simulator and test track) that are traditionally used to gain insights into driver behavior \citep{neale_overview_2005}. NDSs make use of unobtrusive instrumentation (e.g., cameras, sensors) that are typically attached to the primary vehicle of study participants to collect data on driving parameters, such as speed, acceleration, and global positioning system (GPS) location, from on to off ignition \citep{dingus_naturalistic_2014, dingus_100-car_2006}. This aspect of NDSs coupled with the reduced observer effects and lack of special driving instructions enable researchers to study driver behavior with high precision and validity \citep{neale_overview_2005, foss_distracted_2014}. Additionally, rapid advancements in cloud storage and cloud computing technologies have made it considerably easier and cheaper to store and process vast amounts of data \citep{byrne_rise_2018, krumm_practical_2020}. This has propelled the use of NDSs for driver behavior research even further over the past two decades. Reflecting this, the number of search results with the phrase ``naturalistic driving study'' in articles indexed in Google Scholar has increased 600-percent from 678 during 2000--2010 to 4,930 during 2011--2020 \citep{google_google_2021}.

\begin{figure}[!htbp]
\centering
\includegraphics[width = 0.7\linewidth]{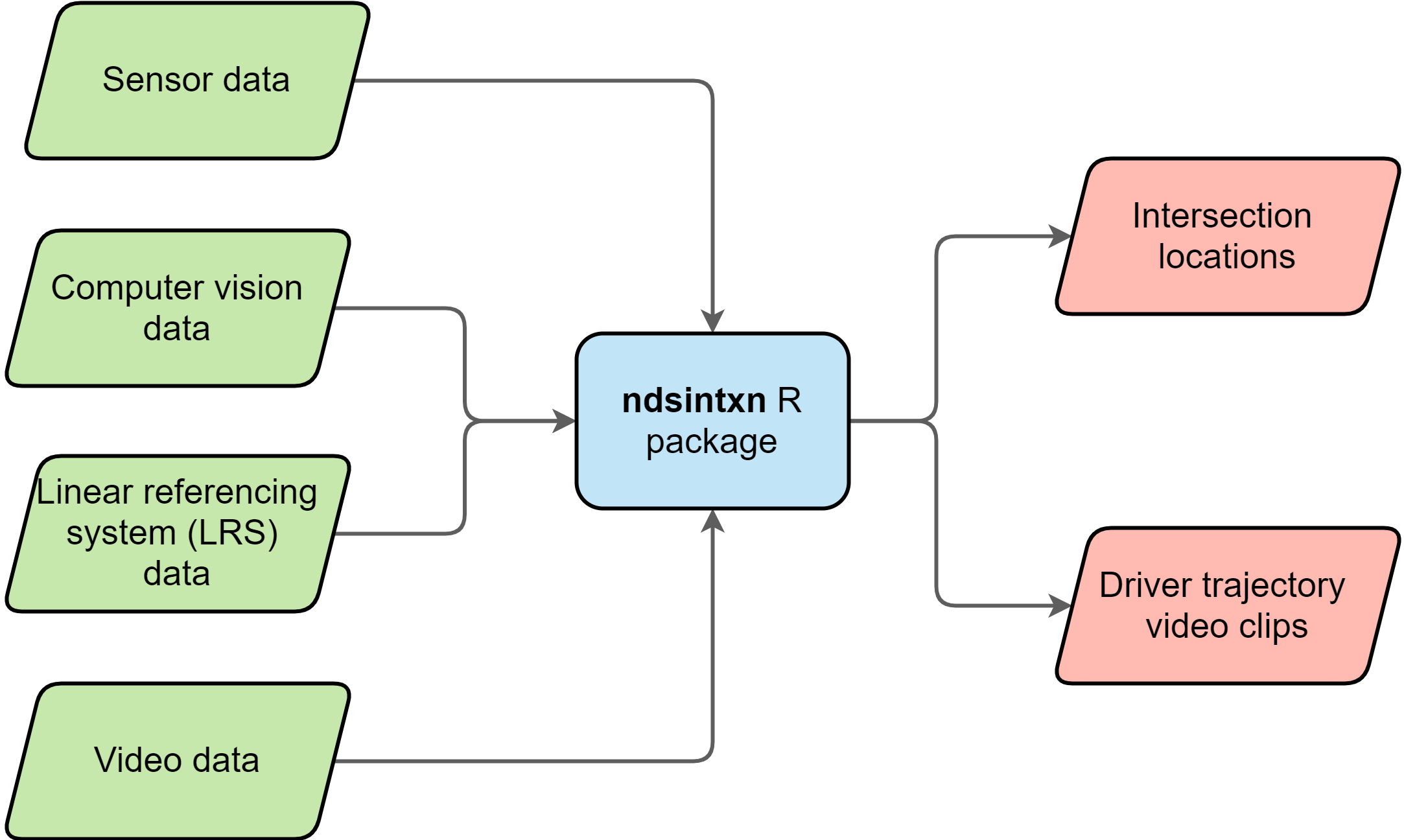}
\caption{High-level overview of the ndsintxn R package}
\label{fig:package-overview}
\end{figure}

One of the many applications of NDSs has been the assessment of typical driver behavior at intersections at a higher level of real-world context than was possible earlier to look for ways to improve safety \citep{wu_driver_2017, xiong_examination_2015}. Identifying strategies to address intersection safety is an active area of research because nearly 25-percent fatalities and about 50-percent injuries in traffic crashes reported each year across the United States are attributed to intersections \citep{national_highway_traffic_safety_administration_fars_2020}. These statistics highlight that the use of NDSs for intersection safety research is only going to grow in the coming future. One common thing among all NDS-based intersection safety research is that they often require specific geographic information on intersection locations as well as video clips of driver trajectories passing through intersections to extract safety-related driver behavior (e.g., driver stopping or turning decision, presence/absence of a lead vehicle or a crossing vehicle, etc.). Intersection identification and trajectory video extraction are complex processes that require merging and processing of data from multiple sources, including video, geographic information system (GIS), street map, and vehicle sensor data. Typically, this process is done with labor-intensive, repetitive, human annotation, constraining data analysis scope due to time, effort, and funding limitations. To address these limitations and improve the data analytic ability of NDSs to comprehensively address driver behavior across geographically diverse intersections, we created a novel R software package called ndsintxn (https://github.com/ashirwad/ndsintxn) that focuses on partially automated extraction of traffic-controlled intersections (e.g., stop signs, traffic signals) to partially automate and streamline this process. The image shown in Figure \ref{fig:package-overview} provides a high-level overview of the ndsintxn package.

To use this package, users need to simply install the package and the additional software tools that include R, RStudio, Google Earth, and FFmpeg. These software tools are open source and are available to download for free. Once all the tools are successfully installed, the package includes functions to extract driver video trajectories at geolocated intersections from inputted data (sensor, video, computer vision detections, and linear referencing systems {[}LRS{]}).

\section{Software Description}
The following sections provide a brief description of different components of the ndsintxn R package, and the inner workings and overall goal of major functions included in the package.

\subsection{Software architecture}
The ndsintxn R package leverages the following services and software tools to generate files that support driver behavior analyses at intersections in NDSs:

\begin{enumerate}
\item
  \textbf{Cloud storage}: NDSs often generate huge amounts of data. The first large-scale NDS called the 100-Car Naturalistic Driving Study generated 6.4 terabytes of data \citep{dingus_100-car_2006}. The recently completed Second Strategic Highway Research Program Naturalistic Driving Study collected 2 petabytes of data \citep{dingus_naturalistic_2014}. Given the large size of the data, it is often stored on cloud-based storage services such as Box, Google Drive, or Amazon Simple Storage Service. Storing data on cloud has many advantages such as scalability, sharing, concurrent data access, data versioning (for optimum reproducibility), etc. Consequently, functions in the ndsintxn package are designed to allow fetching data and pushing results from/to cloud storage services using the respective application programming interfaces.
\item
  \textbf{R and RStudio}: R is a widely used statistical computing software language that facilitates data analysis and visualization. RStudio is an integrated development environment that provides additional features to enhance user experience when working with R as well as provides support for other languages such as Python and C++. The ndsintxn package uses R as the primary computation tool and users are encouraged to use RStudio for an enhanced experience.
\item
  \textbf{Google Earth}: Google Earth is a free software program from Google that allows users to view aerial imagery and street view (if available) for almost any location on Earth. Additionally, the software includes tools for adding placemarks, polygons, paths, etc. for geocoding features of interest. A function in the ndsintxn package exports the list of algorithm-identified intersections to a keyhole markup language (KML) file for manually verifying the correctness of the intersection locations and marking intersection approach legs and entering traffic directions for true intersections. This information is later used for extracting driver trajectories of interest.
\item
  \textbf{Python and FFmpeg}: Python is a general-purpose programing language that is extensively used to support data analysis pipelines, and FFmpeg is an open-source tool for processing video and audio files. The use of command-line interface provided by FFmpeg gets trickier for complex operations, and consequently the ndsintxn package uses FFmpeg via ffmpeg-python \citep{kroening_ffmpeg-python_2021}, a Python package that provides bindings for FFmpeg, to extract driver video clips for the trajectories of interest.
\item
  \textbf{Excel workbook}: A function in the ndsintxn package creates a data entry excel workbook for recording information on pertinent variables to be extracted manually by reviewing driver video clips. To minimize data entry errors, data validation is applied on the cells of the fields to be extracted.
\end{enumerate}

\begin{figure}[!htbp]
\centering
\includegraphics[width = 0.7\linewidth]{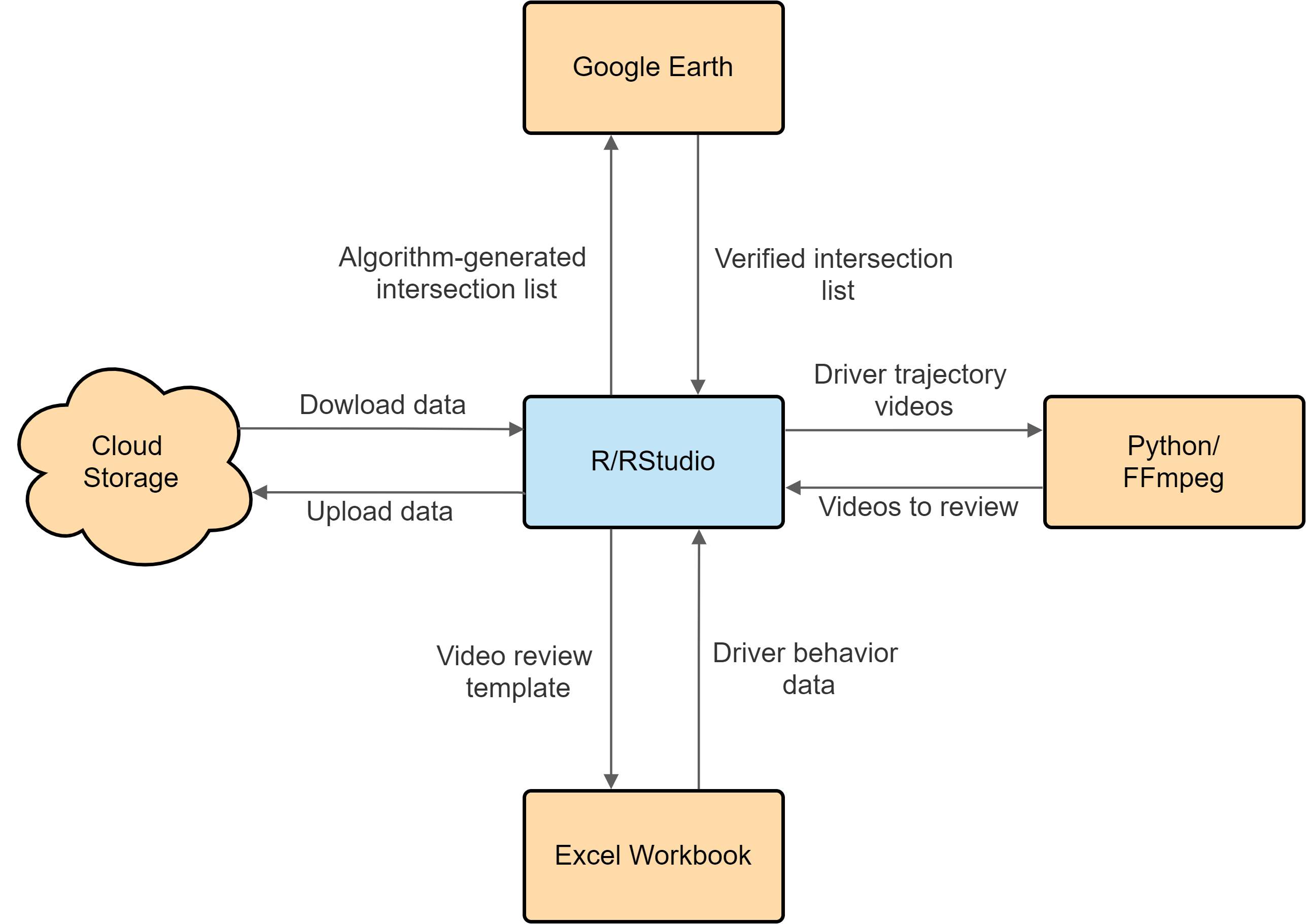}
\caption{Overview of the interactions among software tools and services that support the ndsintxn package}
\label{fig:software-architecture}
\end{figure}

The image shown in Figure \ref{fig:software-architecture} provides an overview of how different components that support the ndsintxn package interact with one another.

\subsection{Software functionalities}

\subsubsection{Data inputs}
In order to use the ndsintxn package, users need the following data inputs: (a) driving GPS coordinates (latitude, longitude), vehicle heading, and vehicle speed collected via sensor systems installed on primary vehicles of study participants (hereafter referred to as ``sensor'' data), (b) timestamps marking when in each driver's data traffic control devices were detected typically obtained from computer vision outputs (hereafter referred to as ``computer vision'' data). This information is obtained by applying computer vision models capable of detecting and localizing traffic element classes to driving video feeds, and (c) LRS file(s) for state(s) where the user wants to locate intersections. This file is generally available on each state's open geographic information system data portal.

\subsubsection{Package functions}
\begin{figure}[!htbp]
\centering
\includegraphics[width = 0.7\linewidth]{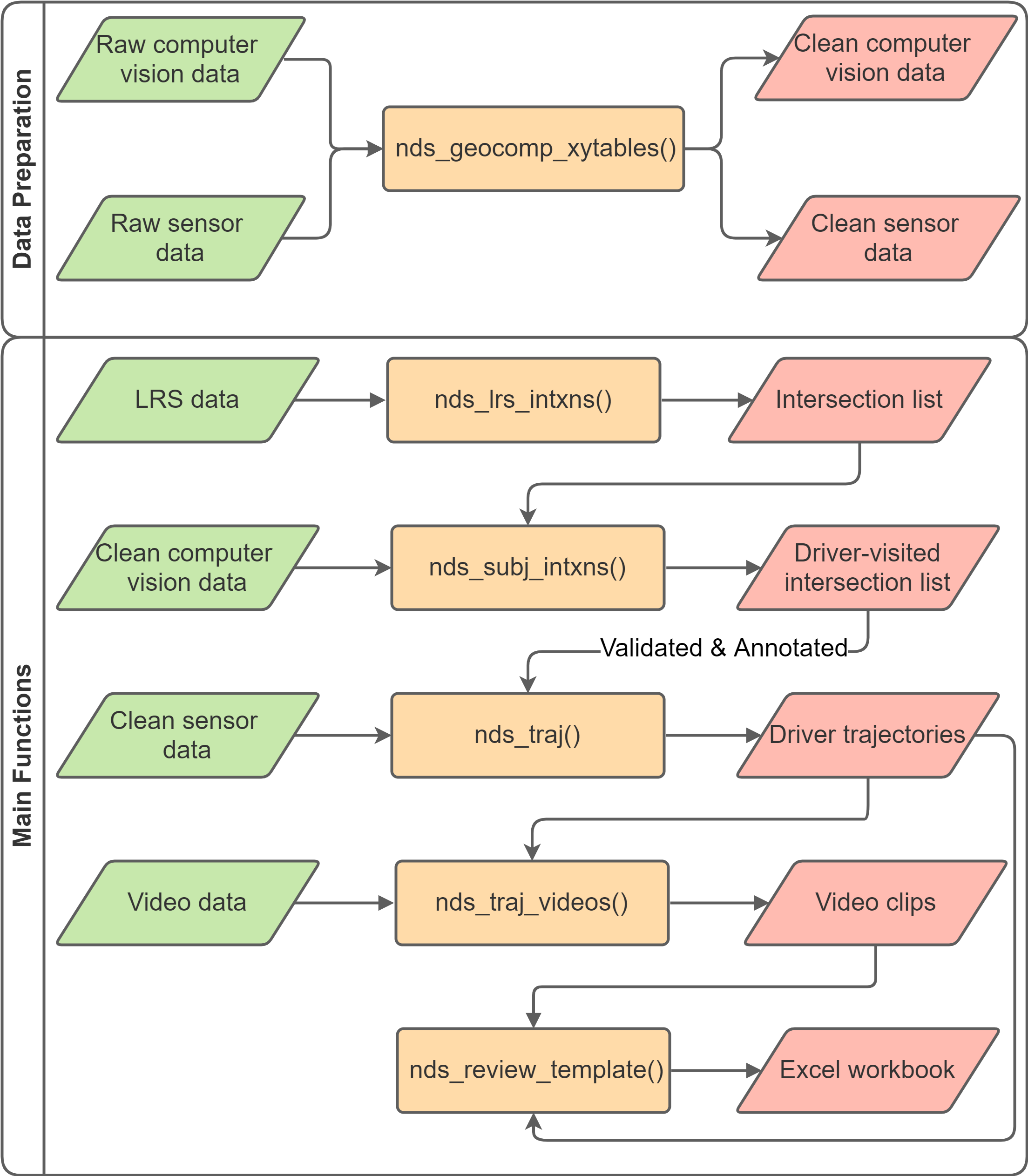}
\caption{Overview of the input and output files for each function in the ndsintxn package}
\label{fig:function-overview}
\end{figure}

There are six functions included in the current version of the ndsintxn package: \linebreak (a) \textbf{nds\_geocomp\_xytables()}: This function processes sensor and computer vision data to create merged data files for geoprocessing operations; (b) \textbf{nds\_lrs\_intxns()}: This function extracts the geographic locations of intersection candidates from an LRS file; (c) \textbf{nds\_subj\_intxns()}: This function extracts the geographic locations of intersection candidates visited by study participants; (d) \textbf{nds\_traj()}: This function extracts study participants' driving trajectories passing through intersections; (e) \textbf{nds\_traj\_videos()}: This function extracts video clips associated with study participants' driving trajectories that need to be manually reviewed to extract pertinent analysis variables; and (f) \textbf{nds\_review\_template()}: This function creates a data entry workbook that a human reviewer will use to record the results of manual video review. All function names are prefixed with the word ``nds'' to avoid namespace ambiguity with functions from other R packages. The image shown in Figure \ref{fig:function-overview} provides an overview of the input and output files for the functions included in the ndsintxn package.

\subsubsection{Implementation details}
The following paragraphs provide details on the implementation of each of the six functions included in the current version of the ndsintxn package.

\begin{enumerate}
\item
  nds\_geocomp\_xytables(): This function processes sensor data and computer vision data for multiple participants to generate two clean files (hereafter referred to as ``clean sensor'' and ``clean computer vision'' files) for the entire study period. The clean files only contain pertinent variables needed for geoprocessing operations. This step reduces the input file size for functions that perform geoprocessing operations and thus allow for better memory management and reduced computation time.
\item
  nds\_lrs\_intxns(): To identify the geographic locations of potential intersection candidates using an LRS file, this function first extracts vertices from the input LRS line features to a point feature class. Next, geometries of the vertices are compared to identify vertices that are spatially coincident with two or more vertices. These vertices are treated as potential intersection candidates using a rule of thumb that three or more spatially coincident points, one point for the end vertex of each approach leg, define an intersection. Finally, the point features corresponding to these vertices are exported to a point feature class representing the geographic locations of potential intersection candidates.
\item
  nds\_subj\_intxns(): To identify the geographic locations of potential intersection candidates visited by study participants, this function makes use of the clean computer vision data generated by the nds\_geocomp\_xytables() function and the intersection list generated by the nds\_lrs\_intxns() function. The clean computer vision data with information on stop sign or signal state detections, depending on the desired intersection type, is first filtered to keep only the last object detection in a sequence of consecutive object detections for each participant-drive combination (stop signs or signal states start getting detected even when a driver is far away from an intersection). Next, the data with information on last object detections is converted to a point feature class to enable geospatial operations. Next, the defined distance (DBSCAN) clustering method with a minimum features per cluster value of two and a search distance value of 100 feet is used to extract clusters from last object detections based on their spatial distribution. The search distance controls two parameters: (a) cluster membership--the specified number of minimum features per cluster must be found within the search distance for cluster membership, and (b) inter-cluster distance--the minimum distance between two individual clusters is set to be equal to the search distance. Next, the centers of the identified clusters are computed and points in the nds\_lrs\_intxns() function-generated intersection list that are closest to the cluster centers are filtered out. These filtered points are then exported to a point feature class where each feature represents a potential driver-visited intersection candidate location. Lastly, this point feature class is exported to a KML file for a manual review of the intersection candidates in Google Earth to separate true and false detections and mark approach legs controlled by a stop sign or a traffic signal for true intersections.
\item
  nds\_traj(): To extract the list of potential driver trajectory candidates passing through the verified intersections, this function makes use of the reviewed data from Google Earth and clean sensor data. The bearings of the lines that mark the directions of entering traffic in the reviewed Google Earth file are first computed and are then transferred to the polygons in the Google Earth file that mark the approach legs of the intersections controlled by a traffic control device (stop sign or a traffic signal) through a spatial join. Next, driver trajectory points that fall within the approach leg polygons are identified and the GPS headings of the trajectory points are compared against the bearings of the entering traffic direction to separate entering and exiting driver trajectories. Next, values in the cumulative distance field of the entering trajectory points are used to identify trajectory points 300 feet upstream and 200 feet downstream of the trajectory point within the intersection approach leg polygons that is closest to the intersection. Lastly, the start and end timings of the entering trajectory candidates are also extracted.
\item
  nds\_traj\_videos(): To extract video clips associated with driver trajectories passing through intersections, this function makes use of driving video feeds and driver trajectory information generated by the nds\_traj() function. The list of dashcam videos that include footage associated with driver trajectories is first extracted by comparing the start and end timings of the videos and trajectories. Next, the identified videos are processed using FFmpeg to create one video clip per trajectory. Additionally, to alert human video reviewers of the approaching intersections at which driver behavior can be captured, a red-colored bounding box is overlaid on top of the video clips. This helps remove ambiguities in what situations to are of interest where multiple stop signs or signals are present in a video clip. The bounding box starts showing up when the driver is 150 feet upstream of the intersection and stops showing when the driver is 50 feet downstream.
\item
  nds\_review\_template()\textbf{:} This function uses information on video clips generated by the nds\_traj\_videos() function and the data on driver trajectory information generated by the nds\_traj() function to create an excel data entry workbook. The workbook contains three kinds of variables: (a) join variables: these variables are pre-populated and are used for merging data sets; (b) helper variables: these variables are also pre-populated and assist human reviewers in locating video clips and facilitate the overall review process; and (c) review variables: these variables are filled out after reviewing video clips and can be customized by a user. Table \ref{tab:review-template-vars} provides a list of variables that are already populated in the standard annotation template.
\end{enumerate}

\begin{table}[!htbp]
\centering
\caption{List of standard variables available in the review template}
\label{tab:review-template-vars}
\begin{tabular}{@{}ll@{}}
\toprule
\multicolumn{2}{@{}l@{}}{\textbf{ID variables}: These variables are used for joining variables from other data sets} \\
stop\_traj\_id & Stop trajectory ID \\
subj & Participant ID \\
drive & Trip number \\
intxn\_id & Intersection ID \\
ref\_time\_utc & Driver arrival time at the stop bar \\ \midrule
\multicolumn{2}{@{}l@{}}{\makecell[tl]{\textbf{Primary participant information}: These variables help reviewers in verifying \\ if the study participant is driving the vehicle}} \\ 
primary\_sub\_age & Age of the primary participant \\
primary\_subj\_gender & Gender of the primary participant \\ \midrule
\multicolumn{2}{@{}l@{}}{\textbf{Helper information}: These variables help facilitate the review process} \\ 
jump\_to\_ref & \makecell[tl]{Number of seconds into the clip at which a driver is 150 \\ feet away from the intersection} \\
video\_url & Video URL \\ \bottomrule
\end{tabular}
\end{table}

\section{Illustrative Example}
In this section, we illustrate the use of the ndsintxn package functions to identify participant-visited stop-controlled intersections and extract video clips for stopping trajectories using data from the NDS conducted at the University of Nebraska Medical Center (UNMC) to examine the influence of type 1 diabetes on stopping behavior at such intersections \citep{barnwal_sugar_2021}. All participants were consented according to institutional policy under UNMC Institutional Review Board \# 462-16-FB. The whole process of identifying intersections and extracting driver trajectory video clips is broken down into five steps: (a) data preparation, (b) automated intersection extraction, (c) human verification of intersection locations, (d) automated driver trajectory video extraction, and (e) manual annotation of driver behavior at intersections. Details on each step are provided in the following paragraphs.

\subsection{Step 1: Data preparation}
The naturalistic driving data was structured such that one sensor and one computer vision file was available for each participant and the data was stored on the Box cloud. As the first step, the nds\_geocomp\_xytables() function was used to process sensor and computer vision files and generate two clean files for the entire study that included information on driving trajectory and stop sign detections for all participants.

\subsection{Step 2: Automated intersection extraction}
\begin{figure}[!htbp]
\centering
\includegraphics{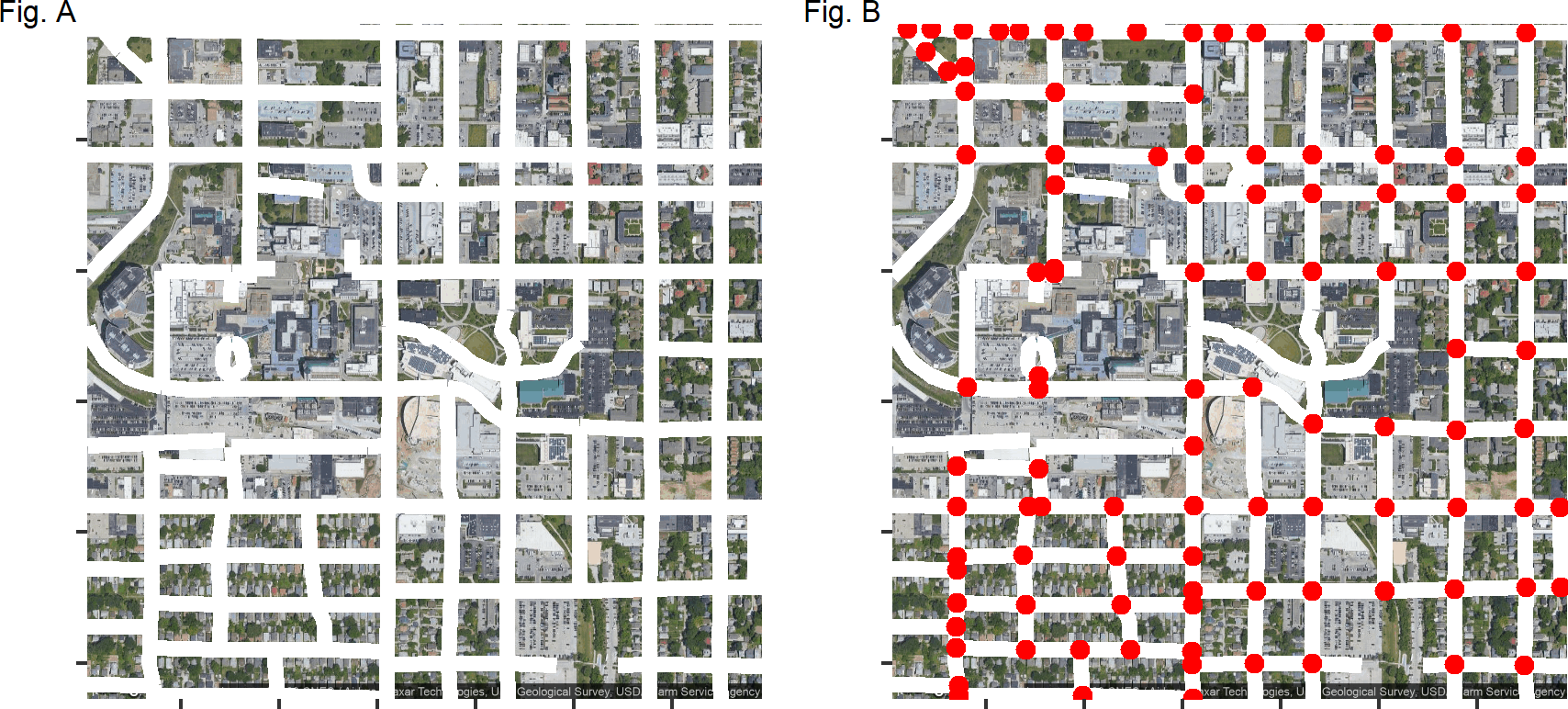}
\caption{Extraction of intersection candidate locations in Nebraska from the LRS file}
\label{fig:lrs-intxn-extraction}
\end{figure}

UNMC study participants were recruited from Omaha, Nebraska, and surrounding areas, and consequently a vast majority of the travel took place within the state of Nebraska. So, Nebraska's LRS file was used as an input in the nds\_lrs\_intxns() function to extract the geographic locations of intersection candidates in Nebraska. The plots in Figure \ref{fig:lrs-intxn-extraction} show the input LRS file with road network highlighted in white color (Fig. A) and the geographic locations of intersection candidates highlighted in red color (Fig. B).

\begin{figure}[!htbp]
\centering
\includegraphics{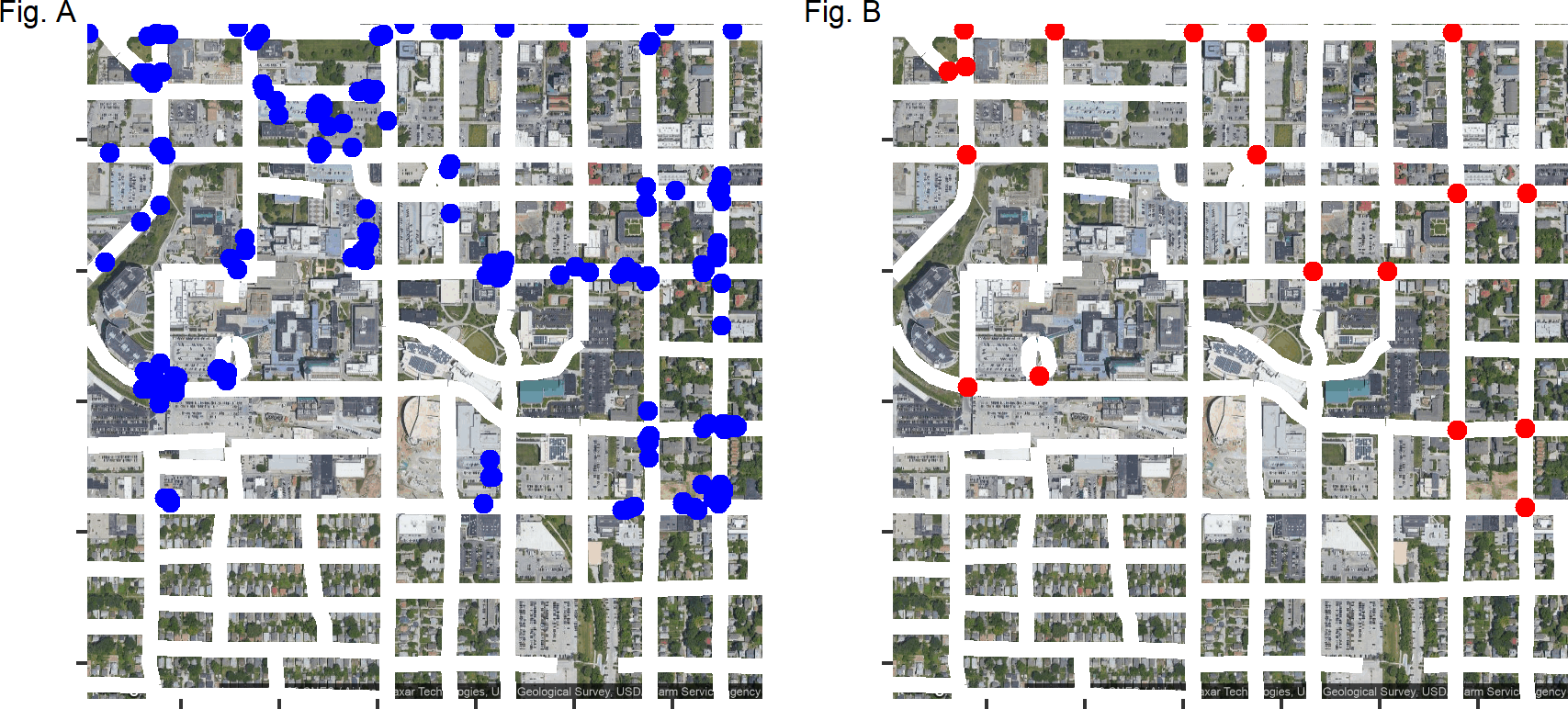}
\caption{Extraction of participant-visited stop-controlled intersection candidate locations in Nebraska using the computer vision data}
\label{fig:subj-visited-intxn-extraction}
\end{figure}

Next, the list of intersection candidates in Nebraska and the computer vision file generated in the data preparation step were used as inputs in the nds\_subj\_intxns() function to identify the geographic locations of stop-controlled intersections visited by the study participants. The plots in Figure \ref{fig:subj-visited-intxn-extraction} show the locations of last stop signs detected during drives undertaken by participants in blue color (Fig. A) and the locations of participant-visited stop-controlled intersection candidates in red color (Fig. B).

\subsection{Step 3: Human verification of intersection locations}
To further verify data accuracy, automatically extracted stop-controlled intersection candidates were manually reviewed in Google Earth to discard false detections. The accuracy rate of the manual review process was about 79-percent. Additionally, for each true intersection, polygons and polylines were drawn to geocode intersection approach legs controlled by stop signs and the direction of entering traffic, respectively. The manual review process took approximately two hours for every 100 candidate locations in the list.

\subsection{Step 4: Automated driver trajectory video extraction}
\begin{figure}[!htbp]
\centering
\includegraphics{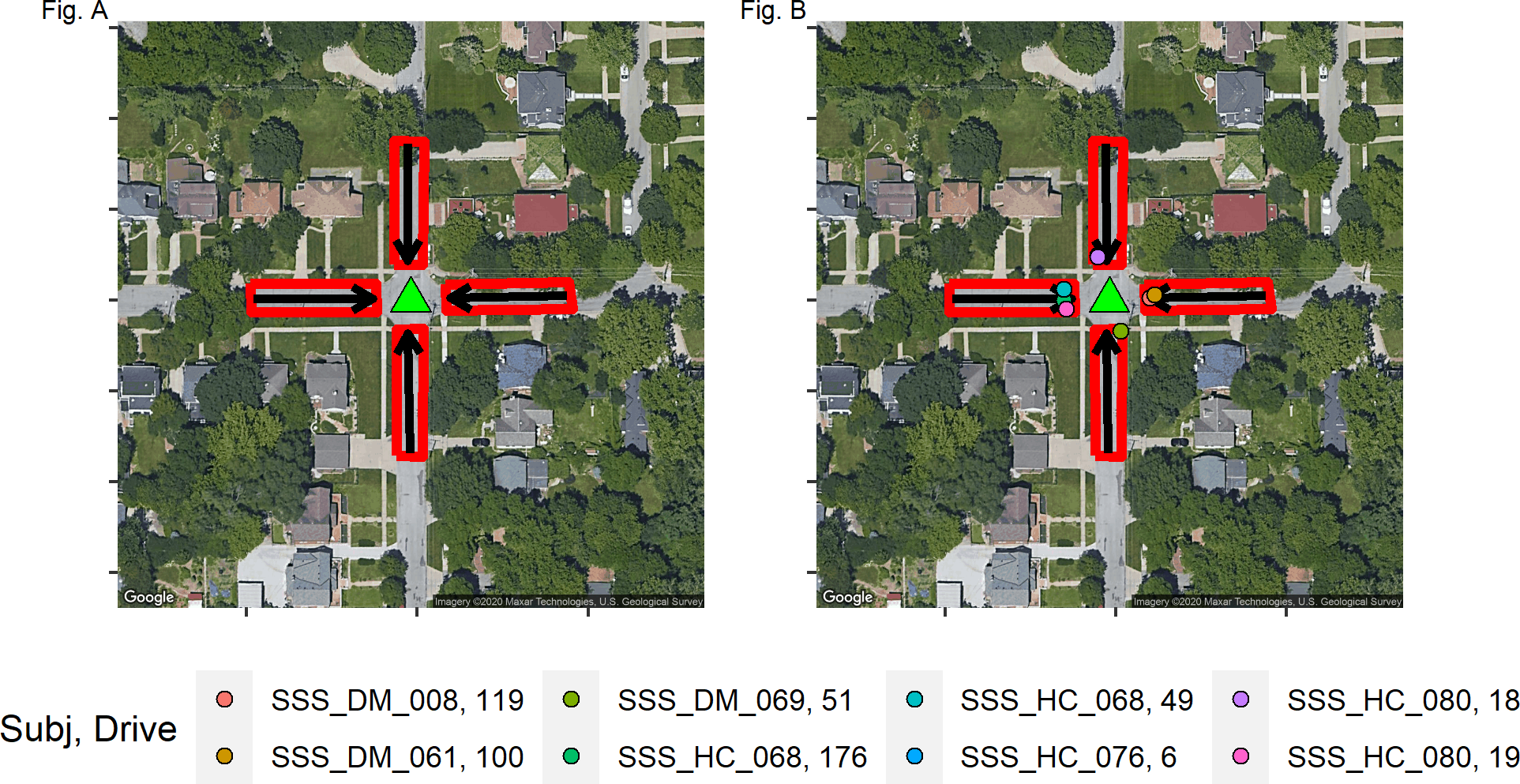}
\caption{Extraction of stop trajectory candidates passing through a stop-controlled intersection}
\label{fig:subj-traj-extraction}
\end{figure}

The reviewed intersection file and the clean sensor data were then used as inputs to the nds\_subj\_traj() function to extract stop trajectory candidates. A unique combination of participant ID, trip number, and stop intersection ID defined a stop trajectory candidate. Fig. A in plots shown in Figure \ref{fig:subj-traj-extraction} show a sample true intersection location (green triangle), stop-controlled approach legs (red rectangles), and the entering traffic direction (black arrows). Stop trajectory candidates passing through this intersection is highlighted with circles colored by trajectory ID in Fig. B.

\begin{figure}[!htbp]
\centering
\includegraphics{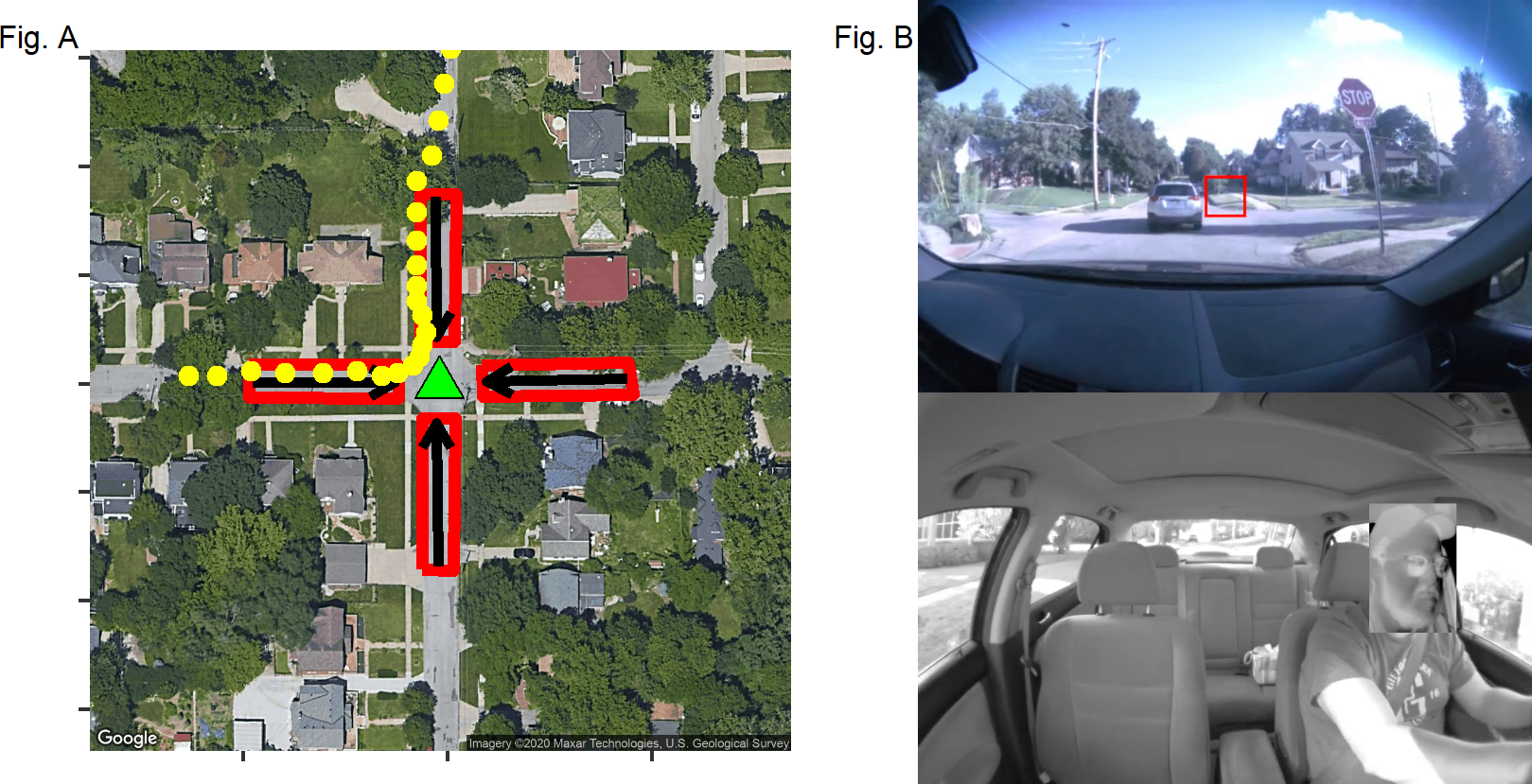}
\caption{Extraction of video clips associated with stop trajectories for manual review}
\label{fig:subj-traj-video-extraction}
\end{figure}

Next, the identified list of stop trajectory candidates and the list of video files were used as inputs to nds\_traj\_videos() function to extract video clips associated with driving trajectories 300 feet upstream and 200 feet downstream of the stop bars. These distances were chosen to ensure that trajectory video clips are long enough to allow for the extraction of all analysis variables. The plots in Figure \ref{fig:subj-traj-video-extraction} show sample trajectory points 300 feet upstream and 200 feet downstream in yellow color (Fig. A) and the video clip associated with the trajectory (Fig. B). The video clip includes footage of both the driving environment and the driver cabin. Additionally, a red square (see the top half of Fig. B) is overlaid in front of the video to alert a human reviewer of an upcoming stop intersection.

\subsection{Step 5: Manual annotation of driver behavior at intersections}
Lastly, the cloud storage information of the extracted video clips and the driver trajectory information were used as inputs to the nds\_review\_template() function to generate a data entry annotation workbook that a human reviewer can use to extract pertinent analysis variables through manual video review. Since we were interested in this analysis if the driver made a safe stop, so the workbook contained custom fields such as type of stop behavior, presence or absence of a lead or a crossing vehicle, etc. in addition to the standard fields shown in Table \ref{tab:review-template-vars}. On average, it took about three hours for a single person to review 100 videos to extract and annotate driver behavior (e.g., driver stopped at stop sign or did not stop).

\section{Software Benefits}
In NDSs, the time it takes to collect the entire driving data is dependent on two things: (a) the pace of participant recruitment, which often happens in phases, and (b) the duration of data collection for each participant. This process generally takes many months, and in some cases can even go up to a several years. So, it is not feasible for researchers to wait for the entire data collection to be over before they can begin driver behavior analyses. Additionally, object detection using computer vision models for the entire driving video feeds require extensive computational power and time. In one of our in-house naturalistic driving projects, we used Fast R-CNN \citep{girshick_fast_2015} on a server with 2 Intel(R) Xeon(R) CPU E5-2630 v4 and 4 GPU: NVIDIA Tesla P-100 with 16 MB computer memory (RAM) each for detecting 32 traffic elements such as vehicles, pedestrians, and traffic signs. The video data for 183,964 driving miles that were collected over a period of one year was 6 terabytes in size and it took about 28 days of continuous running of 4 GPUs to complete the computer vision object detection task. This process was executed using Bash scripts and did not require any human intervention.

To address the issues discussed above, the ndsintxn package allows users to work with whatever driving data have been collected so far in the study. Additionally, users only need to generate computer vision data from the driving video feeds that have been collected so far. The available sensor and computer vision data can then be used with the ndsintxn package to generate a representative sample of geographically dispersed intersections. If more intersections are desired, computer vision data for additional participants can be generated and the process can be repeated to expand the list of intersections. Furthermore, as more driving data become available, the script can be rerun to extract additional driver trajectories to increase sample size. Another major benefit is that for subsequent NDSs, users will already have a list of intersections available. So, they can run the script and check if there are enough trajectories passing through the intersections. If the sample size is enough, users do not need the computer vision data. This cuts down the time significantly and users can focus on extracting analysis variables from the driver video clips. In the illustrative example, all the participants were recruited from Omaha, Nebraska, and surrounding areas, so the list of intersections prepared for this project could also be used in other projects with a similar geographic recruitment area. This strategy can significantly reduce computational needs and data analysis time for newly collected data.

\section{Conclusion}
NDSs are critical for understanding driver behavior in real-world contexts, which are not always well predicted in controlled on-road or simulation environments. We address the large data analytic needs of NDSs to comprehensively address driver behavior across diverse geographic environments via our partially automated algorithm capable of merging rich driver behavior and environmental context information. This lays the platform for developing detailed and informative driver behavior/safety models to inform road safety, roadway design, public policy, etc.

\section{Acknowledgement}
We thank the Mind \& Brain Health Labs at UNMC's Department of Neurological Sciences and Toyota Collaborative Safety Research Center for research and funding resources. We also thank Michelle Nutting for her help annotating and verifying this paper's illustrative data.

\Urlmuskip=0mu plus 1mu\relax
\bibliographystyle{elsarticle-harv}
\clearpage
\bibliography{references}

\end{document}